\documentclass[aps,twocolumn,groupedaddress,amsmath,amssymb]{revtex4-2}
\usepackage[utf8]{inputenc}
\usepackage[T1]{fontenc}
\usepackage{amsmath}
\usepackage{amsfonts}
\usepackage{amssymb}
\usepackage{graphics,graphicx}

\usepackage{graphicx}
\usepackage{subfigure} %colocar figura lado a lado, figura a) e b).
\usepackage{wrapfig} % pacote reponsavel para colocar figura ao lado do texto
\usepackage{epstopdf}
\usepackage{xcolor}
\graphicspath{{figuras/}}

\usepackage[breaklinks=true]{hyperref}
\usepackage{setspace}
\usepackage{graphicx}
\usepackage{color}

\begin{document}

\title{Studies of transport coefficients in charged AdS$_{4}$ black holes on $\kappa$-deformed space}
\author{Fabiano F. Santos}
\email{ fabiano.ffs23@gmail.com}\affiliation{Instituto de F\'isica, Universidade Federal do Rio de Janeiro, 21941-972 Rio de Janeiro, Brazil}
\author{Bruno G. da Costa}
\email{bruno.costa@ifsertao-pe.edu.br}\affiliation{Instituto Federal de Educa\c c\~ao, Ci\^encia e Tecnologia do Sert\~ao Pernambucano, Rua Maria Luiza de Ara\'ujo Gomes Cabral s/n, 56316-686 Petrolina, Pernambuco, Brazil}
\author{Ignacio S. Gomez}
\email{nachosky@fisica.unlp.edu.ar}\affiliation{Instituto de F\'isica, Universidade Federal da Bahia-Campus Universit\'ario de Ondina, 40170-115 Salvador, Bahia, Brazil}

\begin{abstract}
In this work, we study the effect of $\kappa$-deformed space on the thermodynamic quantities, this are find through the holographic renormalization that provide the free energy, which is fundamental to derive the another thermodynamic quantities. For this scenario we consider an charged AdS$_{4}$ black hole for an Einstein-Maxwell model where the derivative quadrivector is replaced by a deformed version inspired in Kaniadakis statistics. Besides, we extract the transport coefficient know as electrical conductivity.
\end{abstract}

\maketitle
%\tableofcontents
\newpage

\section{Introduction}

In recent years the AdS/CFT correspondence has provided a holographic formulation with respect to the field theory in terms of classical gravitational dynamics in a superior spacetime dimension \cite{Maldacena:1997re,Gubser:1998bc,Witten:1998qj,Aharony:1999ti}. The main idea is to obtain a simplified dynamics of the field theory for the effective classical fluid dynamics \cite{Bhattacharyya:2008jc}. In this correspondence, there is a duality between hydrodynamic description and gravitational dynamics. In this way, we can understand aspects of the phase structure of black hole solutions and their stability in terms of the fluid model. In this hydrodynamic regime, we have a set of transport coefficients \cite{Eling:2011ct,Rangamani:2009xk,Aarts:2008yw}.

Some transports coefficients were presented by \cite{Jiang:2017imk,Baggioli:2017ojd,Liu:2018hzo,Li:2018kqp}, which provided universal bounds of the transport coefficients \cite{Hartnoll:2009sz,Sachdev:2011wg}, including the well-know bound ratio of the shear viscosity \cite{Kovtun:2003wp,Kovtun:2004de} that is conjectured on the holographic ``bottom-up'' models. Many of these bounds are violated through some ways 
\cite{Feng:2015oea,Sadeghi:2018vrf} by taking into account the AdS/CFT correspondence in the context of the modified gravity theories \cite{Feng:2015oea,Liu:2018hzo}. In addition, the holographic description has been employed to study black holes from the viewpoint of the information theory within the context of the modified gravity \cite{Jiang:2017imk,Feng:2015oea,Liu:2018hzo,Santos:2020xox,Santos:2020lmb,Santos:2021orr}. These models maintain some of its essential properties such as a second order of the equations of motion, 
as a consequence of a diffeomorphism action that is Lorentz invariant \cite{Heisenberg:2018vsk}.

Thus, motivated by these works we propose new modification by means of a deformed algebraic structure \cite{daCosta:2020mbf,Kaniadakis:2002ie,Kaniadakis,Kaniadakis:2001cqm}, which emerges from a generalization of the Boltzmann-Gibbs entropy. Such algebra has some interesting properties and functions as the $\kappa$-exponential defined by:
\begin{equation}
\exp_\kappa{u} \equiv 
\left( \kappa u + \sqrt{1+\kappa^{2}u^{2}}\right)^{1/\kappa}
=\exp\left[\frac{1}{\kappa}\textrm{arcsinh}(\kappa u)\right],
\end{equation} 
where $\kappa\in\mathbb{R}$, and its inverse function wrote as
\begin{equation}
\ln_\kappa u\equiv\frac{u^{\kappa}-u^{-\kappa}}{2\kappa}=
\frac{1}{\kappa} \textrm{arcsinh}(\kappa\ln u),\quad(u>0).
\end{equation} 
Such functions defined above satisfy some properties as: 
$\exp_\kappa(x)/\exp_\kappa(y)=\exp_\kappa(x\ominus_\kappa y)$, 
$\ln_\kappa(x/y)=\ln_\kappa(x)\ominus_\kappa \ln_\kappa(y)$ 
and $x\ominus_\kappa y\equiv x\sqrt{1+\kappa^{2}y^{2}}-y\sqrt{1+\kappa^{2}x^{2}}$, 
(for more details see \cite{daCosta:2020mbf,Kaniadakis}). 
For the limit $\kappa\to0$, we have that the ordinary exponential and logarithmic functions return to the usual case. A $\kappa$-deformed derivative operator was presented by \cite{Kaniadakis} as
\begin{eqnarray}
\label{eq:q-derivative-dual}
\begin{array}{lll}
  \displaystyle D_{\kappa,u} f(u) 
        &\equiv& \displaystyle \lim_{u'\to u}\frac{f(u')-f(u)}{u'\ominus_\kappa u} 
\\[10pt]
        &=&      \displaystyle\sqrt{1+\kappa^{2}u^{2}} \frac{df(u)}{du}.
 \end{array}
\end{eqnarray}
In fact, the generalization of the Maxwell tensor that comes from $\kappa$-deformed algebra maps the scenario of two horizons  where these horizons are inner and the outer where the first one is a non-physical and the second one is the physical horizon, respectively, where this is a characterize of charged black hole solutions to the Einstein-Maxwell-Dilaton (EMD) equations that are asymptotically AdS at finite temperature and density \cite{Ballon-Bayona:2020xls,Chen:2020ath,Chamblin:1999tk,Chamblin:1999hg}. An important physical motivation to mention is that the dilaton fulfills the IR criterion, that is, to confinement in the dual gauge theory as well as linear Regge trajectories for scalar and tensor glueballs. However, it is known fact and well understood that the deformation of AdS space due to a dilaton field in five-dimensions is dual to the deformation of a CFT in four-dimensions due to a scalar operator. In our case, is clear that the $\kappa$-deformed algebra obeys the same structure, i.e, we have that the $\kappa$-deformation in four-dimensions is dual to the deformation of a of CFT in three-dimensions due to a Kaniadakis  operator \cite{Kaniadakis:2002ie,Kaniadakis,Kaniadakis:2001cqm}. From this perspective, we draw a correspondence between the dilation part of Einstein-Maxwell-Dilaton (EMD) and the $\kappa$-deformation, from the kinematic perspective.

We propose a study of the thermodynamics quantities of charged AdS$_{4}$ black hole through the holographic renormalization 
\cite{Santos:2021orr,Hartnoll:2009sz,Ballon-Bayona:2020xls,Chen:2020ath,Chamblin:1999tk,Chamblin:1999hg}. Such quantities are modified by $\kappa$-deformed algebra, which leads a deformation dilaton-like form as presented by \cite{Ballon-Bayona:2020xls}. However, another interesting quantity is the electrical conductivity that in our prescription provide similar effects to the case of electrical conductivity 
of strange metals having an analogy with the graphene.

The work is organized as follows. 
In Sec.$\sim$\ref{v2} we present the holographic setup to be employed. 
In Sec.$\sim$\ref{v3} we address the issue of finding the thermodynamic quantities. 
In Sec.$\sim$\ref{v4} is devoted to explore the electrical conductivity. 
Finally, in Sec.$\sim$\ref{v5} we  draw the conclusions and outline some perspectives.

%------------------------------------------------------------------%
\section{$\kappa$-Deformed holographic setup}\label{v2}
%------------------------------------------------------------------%

\begin{figure}[!ht]
\begin{center}
\includegraphics[scale=0.6]{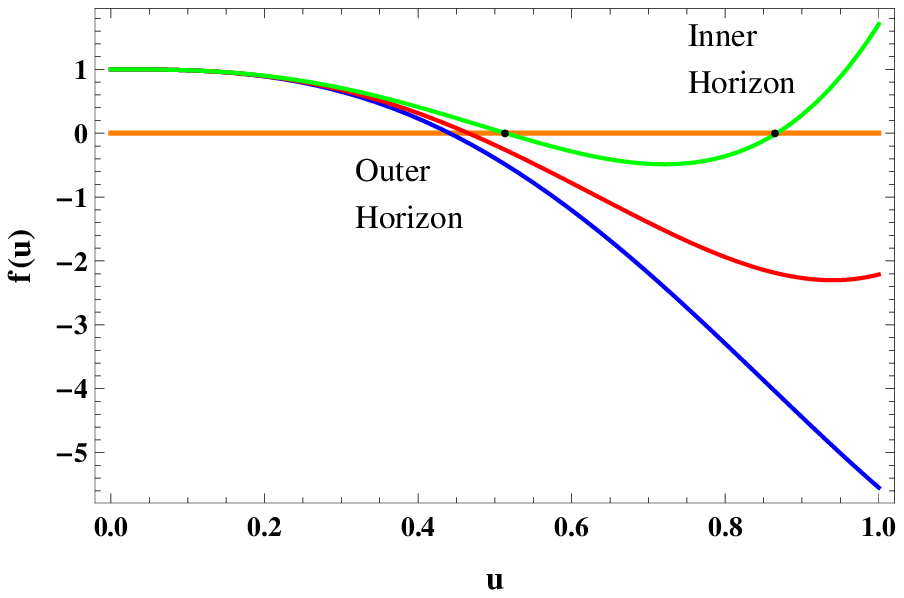}
\includegraphics[scale=0.6]{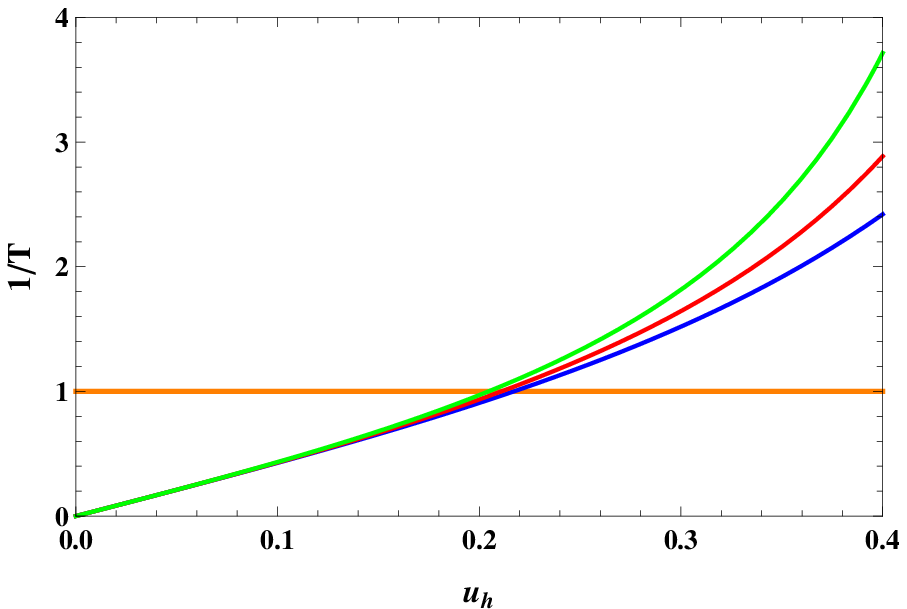}
\caption{The top panel shows the behavior of the equation (\ref{4}) 
$u_{h}=0.4$ and bottom panel shows the temperature equation (\ref{5}) 
both with the values 
$k=2L^{2}$, $\mu=1.2$, $\kappa=0.16$ (curve blue), 
$\mu=1.4$, $\kappa=0.2$ (curve red), 
and $\mu=1.6$, $\kappa=0.25$ (curve green).}
\label{fig:1}
\end{center}
\end{figure}

In this section, we analyze a important example, so-called Reissner-Nordstrom black hole on $\kappa$-deformed space, which corresponds to the dual gravitational description of a quantum field theory QFT at finite temperature $T$ and charge density $\rho$ \cite{Baggioli:2016rdj,Hartnoll:2009sz}. The charge density ingredient can be introduced by the following Einstein-Maxwell action  
\begin{equation}
\label{eq:1}
I=\frac{1}{2k^{2}}\int d^{4}x\sqrt{-g} 
    \left( (R-2\Lambda)-\frac{k^{2}}{4}
        \mathcal{F}_{\mu\nu}\mathcal{F}^{\mu\nu} 
    \right). 
\end{equation}
In order to available the modification in the thermodynamical quantities associated to the Reissner-Nordstrom black hole, we introduce a $\kappa$-deformed Maxwell tensor $\mathcal{F}_{\mu\nu}=D_{\mu}A_{\nu}-D_{\nu}A_{\mu}$. Here $D_{\mu}=\sqrt{1+\kappa^{2}u^{2}}\partial_{\mu}$ 
is the $\kappa$-deformed derivative associated to the Kappa statistics. The charge density of this system can be encoded in the temporal component of the gauge field, which requires a non trivial bulk profile given according to \cite{Baggioli:2016rdj,Jain:2010ip} by $A_{t}(u)$. 
The Einstein-Maxwell field equations can be formally written varying the action (\ref{eq:1}) $\delta I$ as in the usual way

\begin{eqnarray}
\label{2}
&&G_{\mu\nu}+\Lambda g_{\mu\nu}=kT^{em}_{\mu\nu}\\
&&T^{em}_{\mu\nu}
     =\mathcal{F}_{\mu\lambda}\mathcal{F}^{\lambda}_{\nu}
     -\frac{g_{\mu\nu}}{4}\mathcal{F}_{\rho\sigma}\mathcal{F}^{\rho\sigma}.
\end{eqnarray} 
while the Maxwell equation is
\begin{eqnarray}
\tilde{\nabla}^{\mu}\mathcal{F}_{\mu\nu}=0
\end{eqnarray} 
where $\tilde{\nabla}^{\mu}=\sqrt{1+\kappa^{2}u^{2}}\partial^{\mu}$, and as we in the curved spacetime, the solution to the Maxwell equation is the form $A_{t}=\mu-\rho u$. Beyond , we can see that the conservation of the source in Maxwell’s equations impose a constraint in the value of $\kappa=0$, this fact is according to conservation law that corresponds to null energy conditions of the associated stress-tensor. Now, in our prescription we propose a black hole solutions embedded in AdS spacetime with the following metric
\begin{eqnarray}
\label{3}
ds^{2}=\frac{L^{2}}{u^{2}} \left( -f_{\kappa}(u)dt^{2}
        +\frac{du^{2}}{f_{\kappa}(u)}+dx^{2}+dy^{2}\right).
\end{eqnarray}  
Solving the Einstein-Maxwell field equations (\ref{2}) for the metric (\ref{3}), and using $Q^{2}=\rho^{2}u^{4}_{h}$, 
we arrive at
\begin{eqnarray}
\label{4}
f_{\kappa}(u)=1-\left(\frac{u}{u_{h}}\right)^{3}
              +Q^{2}\left(\frac{u}{u_{h}}\right)^{4}
              \left(1+\frac{\kappa^{2}u^{2}}{3}\right).
\end{eqnarray}  
and the temperature is given by
\begin{eqnarray}
\label{5}
T=-\frac{f'_{\kappa}(u_{h})}{4\pi}=
\frac{1}{4\pi}\left(-3\mu^{2}\kappa^{2}u^{3}_{h}
+\frac{3-4\mu^{2}u^{2}_{h}}{u_{h}}\right),
\end{eqnarray}
in which $\mu = \rho u_h$ denotes chemical potential
and $u_h$ is the horizon.
In figure \ref{fig:1}, we have two horizons to the equation (\ref{4}) for the different values $\mu$ and $\kappa$. Such horizons are inner and the outer where the first one is a non-physical and the second one is the physical horizon, respectively. This behavior is according to \cite{Ballon-Bayona:2020xls}. Furthermore, in our case we derive an analytical expressions to the black hole solutions with the $\kappa$-deformation that maps the same behavior of the Einstein-Maxwell-Dilaton. On the other hand, in the usual case of Einstein-Maxwell-Dilaton such behaviors only are possible due numerical computations \cite{Ballon-Bayona:2020xls}.

%------------------------------------------------------%

%------------------------------------------------------%
\section{Thermodynamic quantities}\label{v3}
%------------------------------------------------------%

\begin{figure}[!ht]
\begin{center}
\includegraphics[scale=0.6]{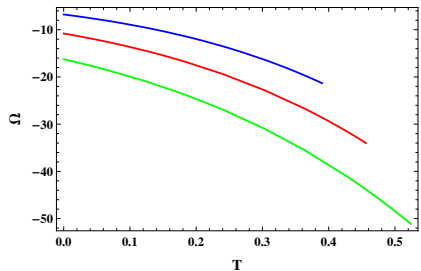}
\caption{The figure shows the behavior of 
the grand potential $u^{1}_{h}$ 
with the values $k=2L^{2}$, 
$\mu=1.2$, $\kappa=0.16$ (curve blue), 
$\mu=1.4$, $\kappa=0.2$ (curve red), 
and $\mu=1.6$, $\kappa=0.25$ (curve green).}
\label{fig:2}
\end{center}
\end{figure}

\begin{figure}[!ht]
\begin{center}
\includegraphics[scale=0.6]{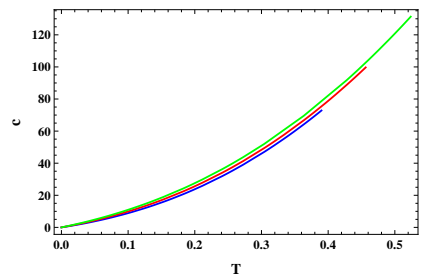}
\caption{The figure shows the behavior of heat capacity, for the $u^{1}_{h}$ 
with the values $k=2L^{2}$, 
$\mu=1.2$, $\kappa=0.16$ (curve blue), 
$\mu=1.4$, $\kappa=0.2$ (curve red), 
and $\mu=1.6$, $\kappa=0.25$ (curve green).}
\label{fig:3}
\end{center}
\end{figure}

Now, we present the thermodynamic of the charged black hole, which is done in the grand canonical ensemble where the idea is extract the potential from the Euclidean action. In this setup, we can to work with a fixed charge density $\rho$, we have that the counterterms 
are necessary to render a finite Euclidean action \cite{Hartnoll:2009sz}. The grand potential $\Omega$ of the quantum field theory (QFT) at finite temperature and charge density with the on-shell euclidean action (\ref{2}). Using $Z=e^{-I_{E}}$ where $\Omega=-T\ln(Z)=-TI_{E,\textrm{on-shell}}$, with these relation is possible to write 
\begin{eqnarray}
\label{6}
\Omega = -\frac{L^{2}V_{2}}{2k^{2}u^{3}_{h}}
         \left[1+2\mu^{2}u^{2}_{h}+\frac{4\mu^{4}
         \kappa^{2}u^{4}_{h}}{3}\right],
\end{eqnarray}
Here $V_2$ is the volume of the unit horizon 2-manifold. By means of the grand potential we can obtain the transport coefficients, as for example, the charge  density ($\rho=-\frac{1}{V_{2}}\frac{\partial\Omega}{\partial\mu}$), the black hole entropy ($S=-\frac{\partial\Omega}{\partial T}$) and the heat capacity ($c=T\frac{\partial S}{\partial T}$). On the other hand, We can to express the horizon in terms of the temperature using the equation (\ref{5}). The behavior of this quantities are show in figure \ref{fig:2}. The figure \ref{fig:3}, show a stable charged black hole, that is $c>0$.

%------------------------------------------------------%
%------------------------------------------------------%

%------------------------------------------------------%
%------------------------------------------------------%

%------------------------------------------------------%
\begin{figure}[!ht]
\begin{center}
\includegraphics[scale=0.6]{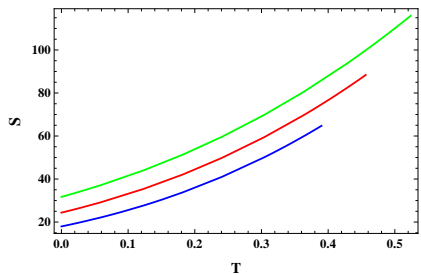}
\includegraphics[scale=0.6]{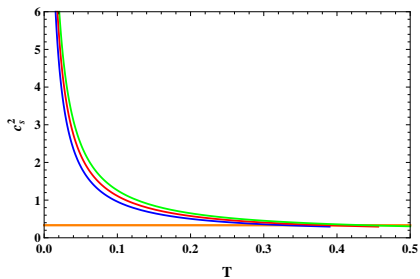}
\caption{The figures show the behavior of 
the entropy and speed of sound for the $u^{1}_{h}$ 
with the values $k=2L^{2}$, 
$\mu=1.2$, $\kappa=0.16$ (curve blue), 
$\mu=1.4$, $\kappa=0.2$ (curve red), 
and $\mu=1.6$, $\kappa=0.25$ (curve green).}
\label{fig:4}
\end{center}
\end{figure}
%------------------------------------------------------%

%------------------------------------------------------%
\begin{figure}[!ht]
\begin{center}
\includegraphics[scale=0.6]{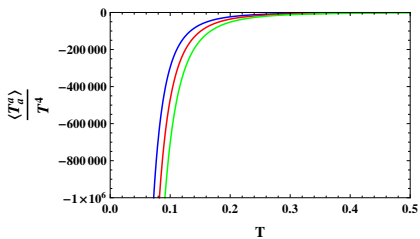}
\caption{The figure shows the behavior of 
the trace of the energy momentum tensor for the $u^{1}_{h}$ 
with the values $k=2L^{2}$, 
$\mu=1.2$, $\kappa=0.16$ (curve blue), 
$\mu=1.4$, $\kappa=0.2$ (curve red), 
and $\mu=1.6$, $\kappa=0.25$ (curve green).}
\label{fig:5}
\end{center}
\end{figure}
%------------------------------------------------------%

%-------------------------------------------------------%
\section{Electrical conductivity}\label{v4}
%-------------------------------------------------------%

\begin{figure}[!ht]
\begin{center}
\includegraphics[width=0.5\linewidth]{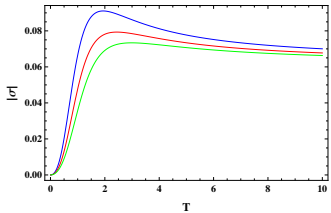}
\caption{
In figure, we have the behavior of 
the electrical conductivity for 
the $u^{1}_{h}$ with the values 
$k=2L^{2}$, 
$\mu=1.2$, $\kappa=1.6$ (curve blue), 
$\mu=1.4$, $\kappa=2.0$ (curve red), and
$\mu=1.6$, $\kappa=2.5$ (curve green), these values are assumed 
to give an acceptable behavior to the electrical conductivity. 
}
\label{fig:6}
\end{center}
\end{figure}

The last quantity of your transport coefficients is the electrical conductivity \cite{Chakrabarti:2010xy}, which is extract using 
the holographic correspondence at the level of response theory, by mean one-point function. These parameters are done in an effective low energy description. Some examples are Well-known examples are shear viscosity $\eta$ \cite{Policastro:2001yc,Policastro:2002se} and 
$DC$ conductivity $\sigma_{DC}$. For the conductivity $\mathcal{O}=J_{r}$, we have that $J_{r}$ is a component of the electric current 
on the level of linear response \cite{Iqbal:2008by,Lucas:2015vna}. Now, following the prescription of \cite{Chakrabarti:2010xy} 
and considering $\mathcal{A}_{t}=\phi(u)e^{-i\omega t+iqx^{2}}$, we can obtain the conductivity in the low frequency limit, as:
\begin{subequations}
\label{7}
\begin{align}
&D_{u}[N(u)D_{u}\phi(u)]+M(u)\phi(u)=0,\\
&N(u)=\sqrt{-g}g^{xx}g^{uu},\\
&M(u)=\sqrt{-g}g^{xx}g^{uu}g^{tt}\mathcal{F}_{ut}\mathcal{F}_{ut}.
\end{align}
\end{subequations}
The electrical conductivity following the works 
\cite{Hartnoll:2008hs,
      Lucas:2015vna,
      Baggioli:2016rdj,
      Chakrabarti:2010xy,
      Hartnoll:2012rj,
      Iqbal:2008by,
      Jain:2010ip,
      Pang:2009wa,
      Jain:2015txa,
      David} 
is given by
\begin{eqnarray}
\sigma
&=& \frac{1}{2k^{2}}\left(\sqrt{\frac{g_{uu}}{g_{tt}}}N(u)\right)_{u=u_{h}}
	\left(\frac{\phi(u_{h})}{\phi(u\to\infty)}\right)^{2}\nonumber\\
&=& \sigma_{H}\left(\frac{Ts}{\epsilon+P}\right)^{2},\label{8}
\end{eqnarray}
where $\sigma_{H}=1/2k^{2}$ is the conductivity evaluated at the horizon. In figure \ref{fig:6}, the electrical conductivity has
an asymptotic behavior at large temperature regime, these limit is due the gap for the values of the chemical potential $\mu$ and $\kappa$. 
This behavior is the analogous has been found in the graphene, for more discussions about this see \cite{David}. Besides, these excitation electrons are associated from the filled valence band into the conduction band, we have that these particle pairs can contribute to the charge density. However, these curves behavior that lie further to the right correspond to systems with higher Fermi energy.

%------------------------------------------------------------%
%------------------------------------------------------------%

\section{Conclusions}\label{v5}
In four-dimensions, we show that the holographic renormalization, equipped with the $\kappa$-derivative provides characteristics similar to 
the Einstein-Maxwell-Dilaton model as presented in \cite{Ballon-Bayona:2020xls}. Such effects are similar to this case, due the fact that 
these $\kappa$-deformed space has a deformation dilaton-like with a quadratic profile. The thermodynamic quantities show a stable 
charged  AdS$_{4}$ black hole. In our dictionary, the trace anomaly of the energy momentum tensor shows that the physical black hole have already merged into a single continuous curve, this behavior indicates that we have a sufficiently large values of $\mu>\mu_{c}$ where $\mu_{c}$ is the critical chemical potential.

Besides, this $\kappa$-holographic transport can mapping the electrical conductivity of strange metals, which is analogous to the graphene, agreement with the usual results founded in CFT$_{3}$ \cite{David}. In fact, these effects are captured by the QFT at low frequencies, 
which is not captured by the Drude model at high frequencies. The plots shown in the figure \ref{4} are not measured at the relativistic Dirac point, but at finite chemical potential, and $\kappa$-algebra, respectively. 

%\section*{Acknowledgment}

I.\ S.\ Gomez acknowledges support received from 
the National Institute of Science and Technology for Complex Systems (INCT-SC), 
and from the Conselho Nacional de Desenvolvimento Cient\'ifico e Tecnol\'ogico 
(CNPq) (at Universidade Federal da Bahia), Brazil.

\end{document}